\documentclass[aps,preprint,showpacs,showkeys]{revtex4-1}
\usepackage{graphicx}
\usepackage{url}
\usepackage{color}
\begin{document}
\title{The electrically charged universe}
\author{Michael~D\"{u}ren}
\email{Michael.Dueren@uni-giessen.de}
\affiliation{II.~Phys.~Inst., Justus-Liebig-Universit\"{a}t Giessen, \\
Heinrich-Buff-Ring 16, 35392 Giessen, Germany}
\date{January 31, 2012}

\begin{abstract}
The paper discusses the possibility of a universe that is not electrically neutral but has a net positive charge. It is claimed that such a universe contains a homogeneous distribution of protons that are not bound to galaxies and fill up the intergalactic space. This proton `gas' charges macroscopic objects like stars and planets, but it does not generate electrostatic or magnetic fields that affect the motion of these bodies significantly. However, the proton gas may contribute significantly to the total dark matter of the universe and its electrostatic potential may contribute to the dark energy and to the expansion of the universe.
\end{abstract}
\pacs{98.80.Bp, 98.80.-k, 95.36.+x}

\keywords{dark matter, dark energy, cosmic rays, intergalactic gas, inflation}
\maketitle

\section{Introduction}
The macroscopic universe consists mainly of electrically neutral objects like for example stars, galaxies and dust and the motion of these objects and apparently also the expansion of the cosmos is governed by gravitational forces. Typically, electric and magnetic forces are much stronger than gravitational forces, but they seem not to play a major role in the dynamics of the cosmos. While strong magnetic forces show up as local sources for synchrotron radiation, strong large-scale electrostatic forces are not observed on the cosmic scale. If two stars close to each other would be electrically charged, electrostatic forces would easily supersede the attractive gravitational force. In this paper the possibility is discussed, that the universe is filed up with a `gas' of positively charged protons that are uniformly distributed in the intergalactic space. Due to symmetry arguments, these protons do not generate electric fields and it is argued that their mass may contribute to the dark matter of the universe and their electrostatic potential may contribute to the dark energy of the universe and to the expansion of the universe.

The paper is structured as follows: Section~\ref{chapbaryon} introduces the possibility of a charge asymmetry of the universe as an analogy to the observed baryon number asymmetry. Section~\ref{chapbig} claims that the intergalactic proton gas is a consequence of the proposed charge asymmetry at the time of the big bang. Section~\ref{chapgas} discusses the effect of the charged gas on the motion of celestial bodies. Section~\ref{chapcosmi} is an excursus to the observed intergalactic protons that appear as cosmic rays.  Section~\ref{chapinvis} argues that the proton gas is hard to detect. Section~\ref{chapneg} excludes the option of a negatively charged universe. Sections~\ref{chapdark}, \ref{chapenergy}, \ref{chapinflation}  and \ref{chaphole} discuss possible relations of the proton gas to dark matter, dark energy, inflation and black holes. Section~\ref{chapconclus} concludes the paper.
\section{Baryon number asymmetry and charge conservation\label{chapbaryon}}

There are, among others, two important conservation laws in physics: the conservation of baryon number and the conservation of charge. Both laws are experimentally well proven and no exception or violation is known~\cite{pdg}. Our observable universe shows a large asymmetry in baryon number: There is much more matter than antimatter, and as the visible baryonic matter is made from protons and neutrons, there are many more baryons than anti-baryons. Due to the conservation law of baryon number, this asymmetry had to be in place already at the time of the big bang or at least shortly afterwards when the universe had temperatures where baryon formation was finished. The exact mechanism of the formation of the matter-antimatter asymmetry is not yet understood. Oscillations of matter and antimatter were observed in the non-baryonic sector~\cite{pdg}, but this alone cannot explain the observed asymmetry in the baryon number. This paper argues that -- as there is a baryon asymmetry in the universe -- it is an admissible hypothesis that there is also an asymmetry in electrical charge. If there was a matter-antimatter asymmetry present in the baryonic sector at the time of baryon formation, it means that there were more quarks than anti-quarks. As quarks have the charges $+2/3~e$ and $-1/3~e$, a flavour symmetric quark plasma would be positively charged. This possible positive charge of the baryonic sector could be exactly compensated by a negative charge in the leptonic sector. But it could also be that it is not exactly compensated. Let us assume that the big bang fireball was positively charged. Due to the law of charge conservation the positive net charge would still be present in today's universe. 

\section{The positively charged universe at the time of the big bang\label{chapbig}}

At the time of the big bang, a mixture of positive, negative and neutral particles was present. Once the fireball cooled down, and most of the antimatter had annihilated with matter and most unstable particles had decayed, the only electrically charged particles left over were protons and electrons and some small portion of their antiparticles, of nuclei, and of unstable particles \cite{unsold}. At the time when the temperature was below the ionisation energy of the hydrogen atom, the negative electron and the positive proton formed neutral hydrogen atoms. Due to Coulomb's law of electrostatics \cite{jackson}, positive and negative charges attract each other over large distances, so that finally all protons would be neutralized by an electron in case there were by chance as many electrons as protons in the universe and the charge asymmetry of the universe were zero. In this case the strong electrostatic forces would diminished and from that on the motion of the universe is governed by the much weaker gravitational forces. The neutral gases start to form neutral dust, stars and galaxies in the early universe \cite{unsold}, \cite{belu}.

In this paper the possibility is discussed that there is a significant charge asymmetry in the universe, which means that the fireball that was generated by the big bang, had a large positive or negative net charge. First the possibility of a large positive net charge is discussed. In this case there must have been a number of protons after the big bang that find no negative partner to neutralize. The electrostatic potential energy of the remaining protons was minimized by two mechanisms:
\begin{enumerate}
\item Due to their repulsion, the remaining charged protons distributed uniformly in the whole universe and
\item the universe expanded using up part of the electrostatic energy of the repelling protons. 
\end{enumerate}
Both mechanisms increased the distance of neighbouring charges. 
The gravitational and electrostatic  potential energies $E_g$ and $E_{el}$ of two protons at distance $r$ are   \cite{misner}, \cite{jackson}
\begin{equation}
E_{g}=-G{M_p^2 \over r}  \qquad  \hbox{and} \quad  E_{el}={1 \over 4\pi\epsilon_0}{e^2 \over r}
\label{eqelg}
\end{equation}
where $G$ is the gravitational constant,  $\epsilon_0$ the electric constant, $e$ the elementary charge and  $M_p$ the proton mass.
The ratio $f$ of gravitational versus electrical energies or forces at any distance for two protons is
\begin{equation}
f:={E_{g} \over E_{el}} =-4\pi\epsilon_0 G \, {{M_p}^2 \over e^2}=-8.09\times 10^{-37}.
\end{equation}
The negative sign expresses that the gravitational and electrostatic forces have opposite effects: gravitational forces slow the expansion of the universe down while electrostatic forces lead to an accelerated expansion. The large value of  $-1/f=1.2\times 10^{36}$ illustrates that already a relatively small number of net charges can have similar effects as large gravitational masses. Therefore it may be important to include the electrostatic potential in models that describe the expansion of the universe. 

\section{The intergalactic proton gas\label{chapgas}}

As the electrostatic repulsion of protons is a factor of $10^{36}$  larger than their gravitational attraction, the protons will distribute basically uniformly in the whole universe and will not mainly gather in the vicinity of galaxies. Electrostatic fields will completely disappear in the universe in a similar way as they disappear inside a metal. If there were remnant electrical fields in the universe due to an inhomogeneous distribution of charges, the protons would move along the field lines until they compensate the charge inhomogeneity. This way, the proton gas provides a natural explanation for the fact that electric forces do not play a role in the motion of celestial bodies, despite the fact that as well stars as galaxies emit large quantities of stellar wind, cosmic rays and jets that have the potential to continuously charge these bodies at high rate.

Due to symmetry arguments, a homogeneous distribution of charges in the universe has no electric field, as there is no direction defined in which the electric field vector could point to. For the same reason, also the gravitational forces generated by the intergalactic proton gas compensate to zero.

Any neutral celestial body of mass $M$, e.g.~a star or a planet, will attract the free protons of the surrounding gas due to its gravitational potential. The protons will `land' on it and charge it, until it reaches a charge $Q$ that generates a repelling electrical force $F_{el}$ which compensates the gravitational force $F_g$:
\begin{equation}
F_{g}=-F_{el} .
\end{equation}
From Coulomb's law, analogous to  Eq.~(\ref{eqelg}), follows
\begin{equation}
G{M M_p \over r^2} =  {1 \over 4\pi\epsilon_0}  {Qe \over r^2}.
\end{equation}
This equation is fulfilled for all distances $r$ if the charge to mass ration $q_M:=Q/M$ of the celestial body has the value
\begin{eqnarray}
q_M  &=& 4\pi\epsilon_0 G{M_p \over e}\label{eqqm}
 \\
&=& 7.75\times 10^{-29}\hbox{ C/kg}
\end{eqnarray}
The conclusion is, that due to the interaction with the surrounding intergalactic proton gas, the charge to mass ratio $Q/M$ of any macroscopic cosmic object is non-zero, independent of the density of the proton gas, and can be calculated from first principles. The charge to mass ratio of a celestial body is very small compared to the one of particles like the proton or electron: 
\begin{equation}
{q_M \over e/M_p}= -f= 8.09\times 10^{-37}; \qquad {q_M \over -e/m_e}=  -4.41\times 10^{-40}.
\end{equation}

What is the effect on the motion of macroscopic objects? The electrostatic force can influence their orbits and has to be added to the gravitational force to obtain the total attractive force $F_{tot}$:
\begin{equation}
F_{tot}=F_{g}+F_{el} = -G{M_1 M_2 \over r^2} + {1 \over 4\pi\epsilon_0}  {Q_1 Q_2 \over r^2} .
\end{equation}
Using equation (\ref{eqqm}) one obtains
\begin{eqnarray}
F_{tot}&=& -G{M_1 M_2 \over r^2} +{1 \over 4\pi\epsilon_0}  q_M^2  {M_1 M_2 \over r^2} \\
&=&-G'{M_1 M_2 \over r^2} 
\end{eqnarray}
with
\begin{eqnarray}
G'&:=&G(1+f )=G\left( 1-8.09\times  10^{-37} \right)= \hbox{const.}
\end{eqnarray}
The electrostatic component that acts on the charged celestial objects is very small. It has the same radial dependence as the gravitational force and has the same effect as if the gravitational constant is reduced by an amount, which is far below its known precision.

If the objects are moving, not only the electrostatic forces play a role but also the magnetic component has to be taken into account. A moving charge $Q_1$ produces a magnetic field $B_{12}$ at the position of a moving charge $Q_2$. The Lorentz force $F_{mag}$ is calculated for the example of two co-moving charges with velocity $v_1$ and $v_2$ according to  \cite{jackson}
\begin{eqnarray}
F_{mag}&=& -Q_2v_2B_{12} \hbox{ with } B_{12} ={\mu_0 \over 4\pi} Q_1{v_1\over (r_1-r_2)^2}\\
&=& -{\mu_0 \over 4\pi} {Q_1Q_2v_1v_2\over (r_1-r_2)^2}\\
&=&F_{el}\, \mu_0\epsilon_0  {v_1v_2}\\
&=&F_{el}\,  {v_1v_2\over c^2}<F_{el}\ll F_{g}
\end{eqnarray}
with $\mu_0 $ being the magnetic constant and $c$ the speed of light. Two moving objects produce a magnetic force on each other that is smaller than their electrical force.

To conclude, the proton gas neutralizes electric fields in the cosmos and it charges up macroscopic celestial bodies, but the resulting electrostatic and magnetic forces between those objects are 36 orders of magnitude below their gravitational attraction.

\section{Cosmic ray protons\label{chapcosmi}}
Cosmic rays are known to consist mainly of individual protons. About 79\% of the primary cosmic ray nucleons are free protons, and most of the rest are helium nuclei~\cite{pdg}.  Only about 1\% are negatively charged electrons~\cite{unsold}. The ratio of antiprotons to protons is $\sim 2 \times 10^{-4}$~\cite{pdg}.  A large part of the cosmic ray protons have large energies, which means that they travel across galaxies and cannot be bound to galaxies by gravitational forces. The gravitational escape velocity for a proton at the position of our solar system to leave the Milky Way is about 550~km/s \cite{pdg} which corresponds to a kinetic energy of 1.6~keV. Therefore it can be assumed that the density of the high-energy cosmic rays that is observed in our solar system is similar to the high-energy part of cosmic rays in the whole universe, provided that the effect of local sources, e.g.~the sun, and the influence of magnetic fields can be neglected. In the energy range above 1~GeV, the intensity of primary nucleons in cosmic rays is about~\cite{pdg}
\begin{equation}
I_{CR} \approx 1.8\times 10^{4} \left( {E\over \hbox{GeV}}\right) ^{-2.7}
{\hbox{nucleons}\over  \hbox{m$^2$s sr GeV}}. 
\end{equation}
The fraction of primary protons is 79\%, yielding an integrated number of  8360 protons/(m$^2$s$\cdot$sr). Using the speed of light, a proton density of about $n_{CR}= 8360\times 4\pi/c=3.5\times 10^{-4}$ protons/m$^3$ and a mass density of the cosmic ray protons with an energy above 1~GeV of $\rho_{CR}= n_{CR}m_p=6\times 10^{-31}$kg/m$^3$ is obtained. This number is small compared to the total baryonic mass density in the universe that is usually quoted as $ 4\times 10^{-28}$kg/m$^3$ \cite{pdg}.

However, the high-energy protons that appear as cosmic rays are not necessarily the `primordial' protons that we talk about in this paper. At the time of the big bang, the primordial protons were in equilibrium with the rest of the matter and they decoupled at a similar time - may be somewhat earlier - from the rest of the fire ball as the photons that form the cosmic microwave background. Therefore it can be assumed that the temperature of the proton gas is low nowadays. A temperature of $T=2.7$~K corresponds to a kinetic energy of $E_{kin}=0.3 $~meV and a proton velocity of $v=260$~m/s. Such energies or velocities are far below what is detected in cosmic ray experiments.

\section{The invisible gas\label{chapinvis}}

As the proton gas has no electron shell, it does not absorb or emit photons in the visible, IR, or UV region and is therefore transparent and invisible. The only remaining cross section of visible photons with free protons is the Compton cross section $\sigma_c$~\cite{compton}:
\begin{equation}
{d\sigma_c\over d\Omega}=\alpha^2r_c^2P(E_\gamma,\theta)^2\left[ P(E_\gamma,\theta)   + P(E\gamma,\theta)^{-1}-1+\cos^2(\theta)  \right]/2
\end{equation}
with the fine structure constant $\alpha$, the scattering angle $\theta$, the Compton wave length of the proton  $r_c=\hbar/M_pc=2.1\times 10^{-16}$~m and 
\begin{equation}
P(E\gamma,\theta)=   {1\over 1+(E_\gamma/M_pc^2)(1-\cos \theta)}\approx 1.
\end{equation}
This yields a total cross section of about
\begin{equation}
{\sigma_c}=8\pi\alpha^2r_c^2/3=0.2~\mu \hbox{b}.
\end{equation}
Assuming a proton gas density of e.g.~$n_p=1.2$ protons/m$^3$ (this number is motivated in section~\ref{chapdark}), the free path length of a photon in view of Compton scattering off the proton gas is $4\times 10^{34}$~m, which means that the proton gas is absolutely transparent for visual photons on a cosmic scale.

Does the proton gas show up experimentally due to interactions with high energetic cosmic rays? Typical hadronic cross sections for $pp$ interactions are on the order of 40~mb~\cite{pdg}, leading to a free path length of $2\times 10^{29}$~m for the given proton gas density. This means that there is no obvious absorption of high energetic cosmic rays in the proposed proton gas, but it has to be further studied if the proton gas modifies the energy spectrum of cosmic ray protons and  if the occasional production of hadronic showers in the proton gas is consistent with experimental data.

It is not obvious how the dynamics and gross properties of the cosmic proton gas should be described. If an individual proton of the gas is displaced, an electrostatic field is generated which is in between the old and new position of the proton. It is a similar effect like the electron-hole creation in a semi-conductor.

The proton gas can also be described as plasma. Using the formula \cite{jackson}
\begin{equation}
\omega_{p}=\sqrt{n_pe^2\over M_p \epsilon_0 }
\label{eq:plasma}
\end{equation}
a plasma frequency of $\omega_{p}=1.5$~Hz is obtained, meaning that the gas is transparent for electromagnetic waves above this frequency.

The proton gas can also be described as electromagnetically polarisable gas that introduces an index of refraction $n \ne 1$. Such an effect would reduce the speed of photons compared to the speed of particles that do not interact electromagnetically, like neutrinos.  The index of refraction is calculated as 
\begin{equation}
n={ \sqrt{\epsilon_r \mu_r}}
\end{equation}
with $\epsilon_r $  and $\mu_r$  being the relative permittivity and permeability. 
The phase and group velocities  $v_p$ and $v_g$ of electromagnetic waves in a medium are given as \cite{jackson}
\begin{eqnarray}
v_{p}={c\over n} \hbox{ ~~and~~ }
v_g={c\over  n+\omega {\partial n\over \partial\omega}}
\end{eqnarray}
Assuming the simple model that the permeability can be neglected,  and that the proton gas acts as Lorentz oscillator, the relative permittivity as a function of the angular frequency $\omega$ is calculated according to the equation \cite{jackson}:
 \begin{equation}
\epsilon_r(\omega)=1+{n_p e^2\over \epsilon_0M_p}{1\over \omega_1^2-\omega^2-i\beta\omega}
\end{equation}
with  $\beta$ being the damping of the oscillation and $\omega_1$ is the effective resonance frequency which is calculated from the undamped angular frequency $\omega_0$ according to 
\begin{eqnarray}
\omega_1^2&=&\omega_0^2-{n_pe^2\over 3\epsilon_0M_p}\\
&=&\omega_0^2-{\omega_p^2 \over 3}
\end{eqnarray}
with $\omega_p$ being the plasma frequency from equation (\ref{eq:plasma}). For frequencies large compared to the plasma frequency  $\omega_p \approx 1.5$ Hz, and by neglecting the damping and the resonance frequency of the free protons ($\beta = 0, ~ \omega_0=0$), the permittivity is obtained as \cite{jackson}
\begin{equation}
\epsilon_r(\omega)=1-{n_p e^2\over \epsilon_0M_p\omega^2}=1-{\omega_p^2\over \omega^2}.
\end{equation}
The corresponding phase (group) velocity is larger (smaller) than the vacuum speed of light by an amount of $\Delta v$ 
\begin{equation}
\Delta v = v_p-c=c-v_g={1\over 2} \left( {\omega_p\over \omega}\right) ^2.
\end{equation}
For visible light (e.g.~$\lambda =500$ nm) a relative reduction of the speed of photons compared to the vacuum speed of light of
\begin{equation}
\Delta v/c=8\times 10^{-32}
\end{equation}
is obtained, which is very small and much smaller than the experimentally known precision of $c$. It is also much smaller than the apparently observed excess of the neutrino speed compared to the speed of light in the OPERA experiment~\cite{opera}.

\section{The negatively charged universe\label{chapneg}}

If the net charge of the universe were negative instead of positive, there would be surplus electrons instead of protons that fill up the intergalactic space. If electrons hit neutral matter, they are bound by atomic forces and produce negatively ionized atoms. Interstellar dust would become negatively ionized. Such a negative ionization would modify spectral lines and would be observed in telescopes. Therefore, a significant negative net charge of the universe is excluded.

\section{Dark matter\label{chapdark}}

The proton gas contributes to the total matter and to the gravitational potential and is therefore part of what is usually called slow dark matter.

In the standard cosmological $\Lambda$CDM model, the universe contains a mass density of cold dark matter of about $ 2.1\times 10^{-27}$~kg/m$^3$~\cite{pdg}. This corresponds to a proton density of about $n_p=1.2$~protons/m$^3$. In principle, a significant fraction of this matter density could come from the intergalactic proton gas. However, the uniform proton gas cannot account for the observed galaxy rotation curves that are evidence for a dark matter component that is gravitationally attracted by galaxies. Nevertheless, the proton gas could still contribute a significant fraction to the dark matter of the universe.

\section{Potential energy of a charged proton gas\label{chapenergy}}

In classical electrodynamics, two protons at a distance $r$ have a positive potential energy $E_{el}$ according to equation (\ref{eqelg}). The potential energy of an individual proton in the vicinity of a number of $N-1$ protons at distance $r_i$ is
\begin{equation}
E(N)={e^2\over 4\pi \epsilon_0}\sum_{i=2}^N{1\over r_i}.
\end{equation}
The potential energy of a proton in a 3-dimensional space with a uniform proton density $n:=N/V$ can be approximated as follows.
It is assumed that each proton $i$ occupies a spherical shell at radius $r_i$, where the shell has the volume $V_0=1/n$ independent of its radius $r_i$. It follows that the $j$-th proton is inside a radius 
\begin{eqnarray}
r_{j}=\sqrt[3]{{3\over 4\pi } (j\cdot V_0 )}=\sqrt[3]{{3\over 4\pi n } j }.
\end{eqnarray}
This proton contributes to the potential energy of the proton at the centre in a range of 
\begin{equation}
{1\over 4\pi \epsilon_0}{e^2\over r_{j-1}}>E_j>{1\over 4\pi \epsilon_0}{e^2\over r_{j}}
\end{equation}
The total potential energy of the central proton in the vicinity of $N-1$ protons is then 
\begin{eqnarray}
{e^2\over 4\pi \epsilon_0}\sum_{j=1}^{N-1}{1\over r_j}>E(N)&>&{e^2\over 4\pi \epsilon_0}\sum_{j=2}^{N}{1\over r_j}\\
&=&{e^2\over \epsilon_0} \sqrt[3]{ n\over  48\pi^2 }\sum_{j=2}^{N}{j^{-1/3}}
\end{eqnarray}
For large $N$ one obtains
\begin{equation}
\sum_{j=0}^{N}{j^{-1/3}}\approx \int^N_0x^{-1/3}dx={3\over 2}N^{2/3}
\end{equation}
i.e.
\begin{eqnarray}
E(N)&\approx&{3\over 4}{e^2\over \epsilon_0} \sqrt[3]{ nN^2\over  6\pi^2 }
\label{eq:tenergy}
\end{eqnarray}
The total electrostatic energy density $\epsilon_{el}$ of the proton gas is obtained by multiplying $E(N)$ with the proton density $n$: 
\begin{eqnarray}
\epsilon_{el}&=&{E_{tot}\over V}={NE(N)\over V}=nE(N)\\
&=&{3e^2n^2\over 4\pi\epsilon_0} \sqrt[3]{{ \pi\over  6 } V^2}
\label{eq:denergy}
\end{eqnarray}
Applying reasonable cosmological numbers to the last equation, e.g.~an energy density that equals the dark energy of the universe: $\epsilon_{el}=1.3\times 10^{-9}$J/m$^3$  \cite{pdg} and a volume of $10^{79}$m$^3$ that corresponds to the Hubble length$^3$, a proton gas density of $n= 10^{-17}$ protons/m$^3$ is obtained.

To conclude on this section, a proton gas of very low density generates easily electrostatic energies that exceed the amount of dark energy in the $\Lambda$CDM model.

\section{Inflation and accelerated expansion of the universe\label{chapinflation}}

In the previous chapters \ref{chapdark} and \ref{chapenergy} it has been shown that the proton gas may contribute significantly to the known values of dark matter and dark energy. However, the values for the dark matter and dark energy densities in the universe are experimentally not obtained by direct observations, but they are derived using the $\Lambda$CDM model which does not include electrostatic effects. Therefore, a quantitative evaluation of the effects of a charged proton gas may not use these values but has to go back to the original observations and apply a cosmological model that includes the electromagnetic effects in curved spacetime. Such a complete evaluation is beyond the scope of this paper, therefore only a qualitative description will be given here.

Let us first discuss why a charged proton gas leads to the expansion of the universe. One could argue that due to symmetry arguments there is no electric field ($E=0$) in the proton gas and according to the equation~\cite{griff}
\begin{equation}
E_{el}={1\over 2} \epsilon_0 E^2 \label{eq:efield}
\end{equation}
there is no field energy ($E_{el}=0$) and no force $F=qE=0$ that drives the expansion. To understand the apparent discrepancy between formulae (\ref{eq:tenergy})  and (\ref{eq:efield}), and to illustrate the effect of curvature in general relativity, we will take an example of  a curved two-dimensional world that is imbedded in a flat 3-dimensional world. Let us consider the molecules in the 2-dimensional surface of a soap bubble. Due to symmetry arguments in the 2-dimensional surface,  there is no net force between the molecules, as the attraction by the neighbouring molecules compensates. Nevertheless, there is a net force and a potential energy that compresses the bubble. This well-known effect is described as effect of the surface tension in three dimensions. The pressure in the bubble depends on the curvature of the 2-dimensional surface; small bubbles with large curvature (i.e.~small radius of curvature) have a large pressure, large bubbles have a small pressure.

In analogy to the surface energy, a curved universe with a charged proton gas can have a vanishing electric field, but nevertheless there is a potential energy that drives the expansion of the universe. The expansion will be strong when the curvature is large and it will be small when the curvature is small. This effect may explain the observed accelerated expansion of the universe. 

While in a pure gravitational model (without a cosmological constant) there are only attractive forces that decelerate the expansion, the electrostatic contribution is always repulsive and leads to an acceleration of the expansion rate of the universe.

The homogeneity, flatness and isotropy of the observed universe is explained in standard cosmological models by the introduction of an inflationary epoch at the time before $10^{-33}$~s after the big bang~\cite{unsold}. In our model, the extreme expansion of the early universe could be caused -- at least partially -- by electrostatic energy. There are two effects that add up here: when the universe was young, its matter and energy density was very high, therefore its curvature was large and, as illustrated in the surface tension model, a large curvature leads to a large pressure due to geometrical effects. Secondly, not only the matter density but also the charge density was large initially. The electrostatic potential of a charge distribution has a $1/R$ divergence towards small $R$. This might explain the extreme expansion at the initial phase of the universe.

 At the time of inflation, baryons have not yet been formed and instead the net charges were quarks or other particles beyond the standard model of particle physics. The phase transition from quarks that have fractional electrical charges to baryons with integer charges may also have played a role in the dynamics of an electrostatically driven expansion of the universe.

\section{Charged black holes\label{chaphole}}

A charged proton gas in the vicinity of a black hole is absorbed by the black hole and charges it. The argument that particles cannot escape from a black hole is only true for particles that are governed by gravitation. In a semi-classical approach it can be assumed that in the presence of electric fields, charged particles can leave a black hole, as electric forces are much stronger than gravitational forces. If the black hole is positively charged, protons may leave it. If there is an angular momentum in the system, the incoming and outgoing protons will produce magnetic fields and the black hole can become a magnetic dipole. Due to Lorentz forces, the protons can leave the black hole only along the magnetic field lines at the poles of the dipole field. The jets and the radio lobes that are observed in some galaxies could be an effect of the ejected proton gas.

\section{Conclusions\label{chapconclus}}
The paper studies the option of an electrically charged universe. In analogy to the observed baryon number asymmetry, a charge asymmetry is assumed to be present in the early universe. It is argued that a positively charged universe leads to a uniform proton gas that fills the intergalactic space. Like the electron gas in a metal it compensates electric fields in the universe. This proton gas is invisible for telescopic observations and has negligible effects on the motion of celestial bodies, but may contribute to the cold dark matter. The possibility of a significant negative charge is excluded. It can be speculated that a proton gas has significant effects on the expansion of the universe, on the amount of dark energy, and on black holes. The paper cannot prove the existence of the proton gas, but the author is not aware of any observation or any scientific argument that rejects this possibility.

If the proposed intergalactic proton gas exists, it cannot be described as a small perturbation of the existing models of the universe because the electrostatic energies have the potential to be much larger than the gravitational effects. Instead, a more serious study of this possibility requires a complete revisit of the cosmological evolution, taking a unified picture of gravity and electromagnetism in curved spacetime into account.

\end{document}